\documentstyle[aps]{revtex}

\begin{document}
\draft
\date{May 7, 2001; replaced October 15, 2001}
\title{Dirac Spinor Waves and Solitons in \\ Anisotropic Taub-NUT Spaces }
\author{Sergiu I. Vacaru}
\address{Dep. of Phys., CSU Fresno, 2345 East San Ramon, \\
Ave.M/S 37 Fresno, Ca 93740-8031, USA \\
{---} \\ E--mail: sergiu$_{-}$vacaru@yahoo.com}

\author{ Florian  Catalin Popa}
\address{Institute of Space Sciences, P.O.Box MG-23, \\
RO 76900, Magurele, Bucharest, Romania\\ 
{---} \\ E--mail: catalin@venus.nipne.ro}

\maketitle
\begin{abstract}
We apply a new general method of anholonomic frames with associated
nonlinear connection structure to construct new classes of exact solutions
of Einstein--Dirac equations in five dimensional (5D) gravity. Such
solutions are parametrized by off--diagonal metrics in coordinate
(holonomic) bases, or, equivalently, by diagonal metrics given with respect
to some anholonomic frames (pentads, or funfbein, satisfing corresponding
constraint relations). We consider two possibilities of generalization of
the Taub NUT metric in order to obtain vacuum solutions of 5D Einsitein
equations with effective renormalization of constants (by higher dimension
anholonomic gravitational interactions) having distinguished anisotropies 
on an angular parameter or on extra dimension coordinate. The constructions 
are extended to solutions describing self--consistent propagations of 3D Dirac
 wave packets in 5D anisotropic Taub NUT spacetimes. We show that such
anisotropic configurations of spinor matter can induce gravitational 3D
solitons being solutions of Kadomtsev--Petviashvili or of sine--Gordon
equations.
\end{abstract}

\vskip0.2cm


\section{Introduction}

Recently one has proposed a new method of construction of exact solutions of
the Einstein equations on (pseudo) Riemannian spaces of three, four and
extra dimensions (in brief, 3D, 4D,...), by applying the formalism of
anholonomic moving frames \cite{vsol}. There were constructed static
solutions for black holes / tori, soliton--dilaton systems and wormhole /
flux tube configurations and for anisotropic generalizations of the Taub NUT
metric \cite{vvt}; all such solutions being, in general, with generic local
anisotropy. The method was elaborated following the geometry of anholonomic
frame (super) bundles and associated nonlinear connections (in brief,
N--connection) \cite{cartan,vsf} which has a number of applications in
generalized Finsler and Lagrange geometry, anholonomic spinor geometry,
(super) gravity and strings with anisotropic (anholonomic) frame structures.

In this paper we restrict our considerations for the 5D Einstein gravity. In
this case the N--connection coefficients are defined by some particular
parametrizations of funfbein, or pentadic, coefficients defining a frame
structure on (pseudo) Riemannian spacetime and describing a gravitational
and matter field dynamics with mixed holonomic (unconstrained) and
anholonomic (constrained) variables. We emphasize that the Einstein gravity
theory in arbitrary dimensions can be equivalently formulated with respect
to both holonomic (coordinate) and anholonomic frames. In the anholonomic
cases the rules of partial and covariant derivation are modified by some
pentad transforms. The point is to find such values of the anholonomic frame
(and associated N--connection) coefficients when the metric is diagonalized
and the Einstein equations are written in a simplified form admitting exact
solutions.

The class of new exact solutions of vacuum Einstein equations describing
anisotropic Taub NUT like spacetimes \cite{vvt} is defined by off--diagonal
metrics if they are given with respect to usual coordinate bases. Such
metrics can be anholonomically transformed into diagonal ones with
coefficients being very similar to the coefficients of the isotropic Taub
NUT solution but having additional dependencies on the 5th coordinate and
angular parameters.

We shall use the term locally anisotropic (spacetime) space (in brief,
anisotropic space) for a (pseudo) Riemannian space provided with an
anholonomic frame structure induced by a procedure of anholonomic
diagonalization of a off--diagonal metric.

The Hawking's \cite{Ha} suggestion that the Euclidean Taub-NUT metric might
give rise to the gravitational analogue of the Yang--Mills instanton holds
true on anisotropic spaces but in this case both the metric and instanton
have some anisotropically renormalized parameters being of higher dimension
gravitational vacuum polarization origin. The anisotropic Euclidean Taub-NUT
metric also satisfies the vacuum Einstein's equations with zero cosmological
constant when the spherical symmetry is deformed, for instance, into
ellipsoidal or even toroidal configuration. Such anisotropic Taub-NUT
metrics can be used for generation of deformations of the space part of the
line element defining an anisotropic modification of the Kaluza-Klein
monopole solutions proposed by Gross and Perry \cite{GP} and Sorkin \cite{So}%
.

In the long-distance limit, neglecting radiation, the relative motion of two
such anisotropic monopoles can be also described by geodesic motions, like
in Ref. \cite{G1}, but these motions are some anholonomic ones with
associated nonlinear connection structure and effective torsion induced by
the anholonomy of the systems of reference used for modelling anisotropies.
The torsion and N--connection corrections vanish if the geometrical objects
are transferred with respect to holonomic (coordinate) frames.

From the mathematical point of view, the new anholonomic geometry of
anisotropic Taub-NUT spaces is also very interesting. In the locally
isotropic Taub-NUT geometry there are four Killing-Yano tensors \cite{GR}.
Three of them form a complex structure realizing the quaternionic algebra
and the Taub-NUT manifold is hyper-K\"{a}hler. In addition to such three
vector-like Killing-Yano tensors, there is a scalar one which exists by
virtue of the metric being of class $D,$ according to Petrov's
classification. Anisotropic deformations of metrics to off--diagonal
components introduce substantial changes in the geometrical picture.
Nevertheless, working with respect to anholonomic frames with associated
nonlinear connection structure the basic properties and relations, even
being anisotropically modified, are preserved and transformed to similar
ones for deformed symmetries \cite{vvt}.

The Schr\"{o}dinger quantum modes in the Euclidean Taub-NUT geometry were
analyzed using algebraic and analytical methods \cite{GR,CV}. The Dirac
equation was studied in such locally isotropic curved backgrounds \cite
{DIRAC}. One of the aims of this paper is to prove that this approach can be
developed as to include into consideration anisotropic Taub-NUT backgrounds
in the context of the standard relativistic gauge-invariant theory \cite
{W,BD} of the Dirac field.

The purpose of the present work is to develop a general $SO(4,1)$
gauge-invariant theory of the Dirac fermions \cite{DKK} which can be
considered for locally anisotropic spaces, for instance, in the external
field of the Kaluza-Klein monopole \cite{DIRAC} which is anisotropically
deformed.

Our goal is also to point out new features of the Einstein theory in higher
dimension spacetime when the locally anisotropic properties, induced by
anholonomic constraints and extra dimension gravity, are emphasized. We
shall analyze such effects by constructing new classes of exact solutions of
the Einstein--Dirac equations defining 3D soliton--spinor configurations
propagating self--consistently in an anisotropic 5D Taub NUT spacetime.

We note that in this paper the 5D spacetime is modeled as a direct time
extension of a 4D Riemannian space provided with a corresponding spinor
structure, i. e. our spinor constructions are not defined by some Clifford
algebra associated to a 5D bilinear form but, for simplicity, they are
considered to be extended from a spinor geometry defined for a 4D Riemannian
space.

We start in Section II with an introduction in the Einstein--Dirac theory
formulated with respect to anholonomic frames with associated nonlinear
connection structure. We write down the Einstein and Dirac equations for
some classes of metric ansatz which can be diagonalized via anholonomic
transforms to a locally anisotropic basis. In Section III we construct exact
solutions for 3D Dirac wave packets propagating in anisotropic backgrounds.
In Section IV we outline two classes of 5D locally anisotropic solutions of
vacuum Einstein equations generalizing the well known Taub NUT metric for
the cases of anisotropic (angular and/or extra dimension) polarizations of
constants and metric coefficients. Section V is devoted to such solutions of
the 5D Einstein -- Dirac equations which are constructed as generalizations
of Taub NUT anisotropic vacuum metrics to configurations with Dirac spinor
energy--momentum source. In Section VI we prove that the Dirac spinor field
in such anisotropic spacetimes can also induce 3D dimensional solitons which
can treated as an anisotropic, in general, non--trivially typological,
soliton--Dirac wave packet configuration propagating self--consistently in
anisotropically deformed Taub NUT spacetime. Finally, in Section VII we
conclude the work.

\section{Einstein--Dirac Equations with Anholono\-mic Variables}

In this Section we introduce an ansatz for pseudo Riemannian off--diagonal
metrics and consider the anholonomic transforms diagonalizing such metrics.
The system of field Einstein equations with the spinor matter
energy--momentum tensor and of Dirac equations are formulated on 5D
pseudo--Riemannian spacetimes constructed as a trivial extension by the time
variable of a 4D Riemannian space (an anisotropic deformation of the Taub
NUT instanton \cite{vvt}).

\subsection{Ansatz for metrics}

We consider a 5D pseudo--Riemannian spacetime of signature $(+,-,-,-,$ $-)$,
with local coordinates 
\[
u^\alpha =(x^i,y^a)=(x^0=t,x^1=r,x^2=\theta ,y^3=s,y^4=p), 
\]
-- or more compactly $u=(x,y)$ -- where the Greek indices are conventionally
split into two subsets $x^i$ and $y^a$ labeled respectively by Latin indices
of type $i,j,k,...=0,1,2$ and $a,b,...=3,4.$ The 5D (pseduo) Riemannian
metric 
\begin{equation}
ds^2=g_{\alpha \beta }du^\alpha du^\beta  \label{metric1}
\end{equation}
is given by a metric ansatz parametrized in the form 
\begin{equation}
{\small g_{\alpha \beta }=\left[ 
\begin{array}{ccccc}
1 & 0 & 0 & 0 & 0 \\ 
0 & g_1+w_1^{\ 2}h_3+n_1^{\ 2}h_4 & w_1w_2h_3+n_1n_2h_4 & w_1h_3 & n_1h_4 \\ 
0 & w_2w_1h_3+n_1n_2h_4 & g_2+w_2^{\ 2}h_3+n_2^{\ 2}h_4 & w_2h_3 & n_2h_4 \\ 
0 & w_1h_3 & w_2h_3 & h_3 & 0 \\ 
0 & n_1h_4 & n_2h_4 & 0 & h_4
\end{array}
\right] ,}  \label{ansatz0}
\end{equation}
where the coefficients are some functions of type 
\begin{eqnarray}
g_{1,2} &=&g_{1,2}(x^1,x^2),h_{3,4}=h_{3,4}(x^1,x^2,s),  \label{qvar} \\
w_{1,2} &=&w_{1,2}(x^1,x^2,s),n_{1,2}=n_{1,2}(x^1,x^2,s).  \nonumber
\end{eqnarray}
Both the inverse matrix (metric) as well the metric (\ref{ansatz0}) is
off--diagonal with respect to the coordinate basis

\begin{equation}
\partial _\alpha \equiv \frac \partial {du^\alpha }=(\partial _i=\frac %
\partial {dx^i},\partial _a=\frac \partial {dy^a})  \label{pder}
\end{equation}
and, its dual basis, 
\begin{equation}
d^\alpha \equiv du^\alpha =(d^i=dx^i,d^a=dy^a).  \label{pdif}
\end{equation}

The metric (\ref{metric1}) with coefficients (\ref{ansatz0}) can be
equivalently rewritten in the diagonal form 
\begin{eqnarray}
\delta s^2 &=&dt^2+g_1\left( x\right) (dx^1)^2+g_2\left( x\right) (dx^2)^2 
\nonumber \\
&{}&+h_3\left( x,s\right) (\delta y^3)^2+h_4\left( x,s\right) (\delta y^4)^2,
\label{dmetric}
\end{eqnarray}
if instead the coordinate bases (\ref{pder}) and (\ref{pdif}) we introduce
the anholonomic frames (anisotropic bases) 
\begin{equation}
{\delta }_\alpha \equiv \frac \delta {du^\alpha }=(\delta _i=\partial
_i-N_i^b(u)\ \partial _b,\partial _a=\frac \partial {dy^a})  \label{dder}
\end{equation}
and 
\begin{equation}
\delta ^\alpha \equiv \delta u^\alpha =(\delta ^i=dx^i,\delta
^a=dy^a+N_k^a(u)\ dx^k)  \label{ddif}
\end{equation}
where the $N$--coefficients are parametrized 
\[
N_0^a=0,\ N_{1,2}^3=w_{1,2}\mbox{ and }N_{1,2}^4=n_{1,2}
\]
and define the associated nonlinear connection (N--connection) structure,
see details in Refs \cite{cartan,vsol,vvt,vsf}. The anisotropic frames (\ref
{dder}) and (\ref{ddif}) are anholonomic because, in general, they satisfy
some anholonomy relations, 
\[
\delta _\alpha \delta _\beta -\delta _\beta \delta _\alpha =W_{\alpha \beta
}^\gamma \delta _\gamma ,
\]
with nontrivial anholonomy coefficients 
\begin{eqnarray}
W_{ij}^k &=&0,W_{aj}^k=0,W_{ia}^k=0,W_{ab}^k=0,W_{ab}^c=0,
\label{anholonomy} \\
W_{ij}^a &=&-\Omega _{ij}^a,W_{bj}^a=-\partial _bN_j^a,W_{ia}^b=\partial
_aN_j^b,  \nonumber
\end{eqnarray}
where 
\[
\Omega _{ij}^a=\delta _jN_i^a-\delta _iN_j^a
\]
is the N--connection curvature. Conventionally, the N--coefficients
decompose the spacetime variables (tensors, spinors and connections) into
sets of mixed holonomic--anholonomic variables (coordinates) provided
respectively with 'holonomic' indices of type $i,j,k,...$ and with
'anholonomic' indices of type $a,b,c,...$. Tensors, metrics and linear
connections with coefficients defined with respect to anisotropic frames (%
\ref{dder}) and (\ref{ddif}) are distinguished (d) by N--coefficients into
holonomic and anholonomic subsets and called, in brief, d--tensors,
d--metrics and d--connections.

\subsection{Einstein equations with anholonomic variables}

The metric (\ref{metric1}) with coefficients (\ref{ansatz0}) (equivalently,
the d--metric (\ref{dmetric})) is assumed to solve the 5D Einstein equations 
\begin{equation}
R_{\alpha \beta }-\frac 12g_{\alpha \beta }R=\kappa \Upsilon _{\alpha \beta
},  \label{5einstein}
\end{equation}
where $\kappa $ and $\Upsilon _{\alpha \beta }$ are respectively the
coupling constant and the energy--momentum tensor.

The nontrivial components of the Ricci tensor for the metric (\ref{metric1})
with coefficients (\ref{ansatz0}) (equivalently, the d--metric (\ref{dmetric}%
)) are 
\begin{eqnarray}
R_1^1 =R_2^2 &=& -\frac 1{2g_1g_2} [ g_2^{\bullet \bullet }-\frac{%
g_1^{\bullet }g_2^{\bullet }}{2g_1}-\frac{(g_2^{\bullet })^2}{2g_2} 
\nonumber \\
&{}& +g_1^{^{\prime \prime }}- \frac{g_1^{^{\prime }}g_2^{^{\prime }}}{2g_2}-%
\frac{(g_1^{^{\prime }})^2}{2g_1}] ,  \label{ricci1a} \\
R_3^3 &=&R_4^4=-\frac \beta {2h_3h_4},  \label{ricci1b} \\
&{}&  \nonumber \\
R_{31} &=&-w_1\frac \beta {2h_4}-\frac{\alpha _1}{2h_4},  \label{ricci1c} \\
R_{32}&=& -w_2\frac \beta {2h_4}-\frac{\alpha _2}{2h_4},  \nonumber \\
&{}&  \nonumber \\
R_{41} &=&-\frac{h_4}{2h_3}\left[ n_1^{**}+\gamma n_1^{*}\right] ,
\label{ricci1d} \\
R_{42} &=& - \frac{h_4}{2h_3}\left[ n_2^{**}+\gamma n_2^{*}\right] , 
\nonumber
\end{eqnarray}
where, for simplicity, the partial derivatives are denoted $h^{\bullet
}=\partial h/\partial x^1,f^{\prime }=\partial f/\partial x^2$ and $%
f^{*}=\partial f/\partial s.$

The scalar curvature is computed 
\[
R=2\left( R_1^1+R_3^3\right) . 
\]

In result of the obtained equalities for some Ricci and Einstein tensor
components, we conclude that for the metric ansatz (\ref{ansatz0}) the
Einstein equations with matter sources are compatible if the coefficients of
the energy--momentum d--tensor give with respect to anholonomic bases (\ref
{dder}) and (\ref{ddif}) satisfy the conditions 
\begin{equation}
\Upsilon _0^0=\Upsilon _1^1+\Upsilon _3^3,\Upsilon _1^1=\Upsilon
_2^2=\Upsilon _1,\Upsilon _3^3=\Upsilon _4^4=\Upsilon _3,  \label{spinorem}
\end{equation}
and could be written in the form 
\begin{eqnarray}
R_1^1 &=&-\kappa \Upsilon _3,  \label{einsteq2a} \\
R_3^3 &=&-\kappa \Upsilon _1,  \label{einsteq2b} \\
R_{3\widehat{i}} &=&\kappa \Upsilon _{3\widehat{i}},  \label{einsteq2c} \\
R_{4\widehat{i}} &=&\kappa \Upsilon _{4\widehat{i}},  \label{einsteq2d}
\end{eqnarray}
where $\widehat{i}=1,2$ and the left parts are given by the components of
the Ricci tensor (\ref{ricci1a})-(\ref{ricci1d}).

The Einstein equations (\ref{5einstein}), equivalently (\ref{einsteq2a})--(%
\ref{einsteq2d}), reduce to this system of second order partial derivation
equations: 
\begin{eqnarray}
g_2^{\bullet \bullet }-\frac{g_1^{\bullet }g_2^{\bullet }}{2g_1}-\frac{%
(g_2^{\bullet })^2}{2g_2}+ &{}&  \nonumber \\
g_1^{^{\prime \prime }}-\frac{g_1^{^{\prime }}g_2^{^{\prime }}}{2g_2}-\frac{%
(g_1^{^{\prime }})^2}{2g_1} &=&-2g_1g_2\Upsilon _3,  \label{einsteq3a} \\
h_4^{**}-\frac{(h_4^{*})^2}{2h_4}-\frac{h_4^{*}h_3^{*}}{2h_3}
&=&-2h_3h_4\Upsilon _1,  \label{einsteq3b} \\
\beta w_i+\alpha _i &=&-2h_4\kappa \Upsilon _{3i},  \label{einsteq3c} \\
n_i^{**}+\gamma n_i^{*} &=&-\frac{2h_3}{h_4}\kappa \Upsilon _{4i},
\label{einsteq3d}
\end{eqnarray}
where 
\begin{eqnarray}
\alpha _1 &=&{h_4^{*}}^{\bullet }-\frac{{h_4^{*}}}2\left( \frac{h_3^{\bullet
}}{h_3}+\frac{h_4^{\bullet }}{h_4}\right) ,  \label{alpha1} \\
\alpha _2 &=&{h_4^{*}}^{\prime }-\frac{{h_4^{*}}}2\left( \frac{h_3^{\prime }%
}{h_3}+\frac{h_4^{\prime }}{h_4}\right) ,  \label{alpha2} \\
\beta &=&h_4^{**}-\frac{(h_4^{*})^2}{2h_4}-\frac{h_4^{*}h_3^{*}}{2h_3},
\label{beta} \\
\gamma &=&\frac 32\frac{h_4}{h_4}^{*}-\frac{h_3}{h_3}^{*},  \label{gamma}
\end{eqnarray}
and the partial derivatives are denoted, for instance, 
\begin{eqnarray*}
g_2^{\bullet } &=&\partial g_2/\partial x^1=\partial g_2/\partial
r,g_1^{^{\prime }}=\partial g_1/\partial x^2=\partial g_1/\theta , \\
h_3^{*} &=&\partial h_3/\partial s=\partial h_3/\partial \varphi \ (%
\mbox{or
}\partial h_3/\partial y^4,\mbox{ for }s=y^4).
\end{eqnarray*}

\subsection{Dirac equations in anisotropic spacetimes}

The problem of definition of spinors in locally anisotropic spaces and in
spaces with higher order anisotropy was solved in Refs. \cite{vsf}. In this
paper we consider locally anisotropic Dirac spinors given with respect to
anholonomic frames with associated N--connection structure on a 5D (pseudo)
Riemannian space $V^{(1,2,2)}$ constructed by a direct time extension of a
4D Riemannian space with two holonomic and two anholonomic variables.

Having an anisotropic d--metric 
\begin{eqnarray*}
g_{\alpha \beta }(u) &=&(g_{ij}(u),h_{ab}(u))=(1,g_{\widehat{i}}(u),h_a(u)),
\\
\widehat{i} &=&1,2;i=0,1,2;a=3,4,
\end{eqnarray*}
defined with respect to an anholonomic basis (\ref{dder}) we can easily
define the funfbein (pentad) fields 
\begin{eqnarray}
f_{\underline{\mu }} &=&f_{\underline{\mu }}^\mu \delta _\mu =\{f_{%
\underline{i}}=f_{\underline{i}}^i\delta _i,f_{\underline{a}}=f_{\underline{a%
}}^a\partial _a\},  \label{pentad1} \\
\ f^{\underline{\mu }} &=&f_\mu ^{\underline{\mu }}\delta ^\mu =\{f^{%
\underline{i}}=f_i^{\underline{i}}d^i,,f^{\underline{a}}=f_a^{\underline{a}%
}\delta ^a\}  \nonumber
\end{eqnarray}
satisfying the conditions 
\begin{eqnarray*}
g_{ij} &=&f_i^{\underline{i}}f_j^{\underline{j}}g_{\underline{i}\underline{j}%
}\mbox{ and }h_{ab}=f_a^{\underline{a}}f_b^{\underline{b}}h_{\underline{a}%
\underline{b}}, \\
g_{\underline{i}\underline{j}} &=&diag[1,-1-1]\mbox{ and }h_{\underline{a}%
\underline{b}}=diag[-1,-1].
\end{eqnarray*}
For a diagonal d-metric of type (\ref{dmetric}) we have 
\[
f_i^{\underline{i}}=\sqrt{\left| g_i\right| }\delta _i^{\underline{i}}%
\mbox{
and }f_a^{\underline{a}}=\sqrt{\left| h_a\right| }\delta _a^{\underline{a}}, 
\]
where $\delta _i^{\underline{i}}$ and $\delta _a^{\underline{a}}$ are
Kronecker's symbols.

We can also introduce the corresponding funfbein which are related with
the off--diagonal metric ansatz (\ref{ansatz0}) for $g_{\alpha \beta },$%
\begin{equation}
e_{\underline{\mu }}=e_{\underline{\mu }}^\mu \partial _\mu \mbox{ and }e^{%
\underline{\mu }}=e_\mu ^{\underline{\mu }}\partial ^\mu   \label{pentad2}
\end{equation}
satisfying the conditions 
\begin{eqnarray*}
g_{\alpha \beta } &=&e_\alpha ^{\underline{\alpha }}e_\beta ^{\underline{%
\beta }}g_{\underline{\alpha }\underline{\beta }}\mbox{ for }g_{\underline{%
\alpha }\underline{\beta }}=diag[1,-1,-1,-1,-1], \\
e_\alpha ^{\underline{\alpha }}e_{\underline{\alpha }}^\mu  &=&\delta
_\alpha ^\mu \mbox{ and }e_\alpha ^{\underline{\alpha }}e_{\underline{\mu }%
}^\alpha =\delta _{\underline{\mu }}^{\underline{\alpha }}.
\end{eqnarray*}

The Dirac spinor fields on locally anisotropic deformations of Taub NUT
spaces, 
\[
\Psi \left( u\right) =[\Psi ^{\overline{\alpha }}\left( u\right) ]=[\psi ^{%
\widehat{I}}\left( u\right) ,\chi _{\widehat{I}}\left( u\right) ], 
\]
where $\widehat{I}=0,1,$ are defined with respect to the 4D Euclidean
tangent subspace belonging the tangent space to $V^{(1,2,2)}.$ The $4\times
4 $ dimensional gamma matrices $\gamma ^{\underline{\alpha }^{\prime
}}=[\gamma ^{\underline{1}^{\prime }},\gamma ^{\underline{2}^{\prime
}},\gamma ^{\underline{3}^{\prime }},\gamma ^{\underline{4}^{\prime }}]$ are
defined as to satisfy the relation 
\begin{equation}
\left\{ \gamma ^{\underline{\alpha }^{\prime }},\,\gamma ^{\underline{\beta }%
^{\prime }}\right\} =2g^{\underline{\alpha }\underline{^{\prime }\beta }%
^{\prime }},  \label{gammarel}
\end{equation}
where $\left\{ \gamma ^{\underline{\alpha }^{\prime }}\,\gamma ^{\underline{%
\beta }^{\prime }}\right\} $ is a symmetric commutator, $g^{\underline{%
\alpha }\underline{^{\prime }\beta }^{\prime }}=(-1,-1,-1,-1),$ which
generates a Clifford algebra distinguished on two holonomic and two
anholonomic directions (hereafter the primed indices will run values on the
Euclidean and/or Riemannian, 4D component of the 5D pseudo--Riemannian
spacetime). In order to extend the (\ref{gammarel}) relations for unprimed
indices $\alpha ,\beta ...$ we conventionally complete the set of primed
gamma matrices with a matrix $\gamma ^{\underline{0}},$ i. .e. write $\gamma
^{\underline{\alpha }}=[\gamma ^{\underline{0}},\gamma ^{\underline{1}%
},\gamma ^{\underline{2}},\gamma ^{\underline{3}},\gamma ^{\underline{4}}]$
when 
\[
\left\{ \gamma ^{\underline{\alpha }},\,\gamma ^{\underline{\beta }}\right\}
=2g^{\underline{\alpha }\underline{\beta }}. 
\]

The coefficients of gamma matrices can be computed with respect to
coordinate bases (\ref{pder}) or with respect to anholonomic bases (\ref
{dder}) by using respectively the funfbein coefficients (\ref{pentad1}) and (%
\ref{pentad2}), 
\[
\gamma ^\alpha (u)=e_{\underline{\alpha }}^\alpha (u)\gamma ^{\underline{%
\alpha }}\mbox{ and }\widehat{\gamma }^\beta (u)=f_{\underline{\beta }%
}^\beta (u)\gamma ^{\underline{\beta }},
\]
were by $\gamma ^\alpha (u)$ we denote the curved spacetime gamma matrices
and by $\widehat{\gamma }^\beta (u)$ we denote the gamma matrices adapted to
the N--connection structure..

The covariant derivation of Dirac spinor field $\Psi \left( u\right) ,$ $%
\nabla _\alpha \Psi ,$ can be defined with  respect to a pentad
decomposition of the off--diagonal metric (\ref{ansatz0})

\begin{equation}
\nabla _\alpha \Psi =\left[ \partial _\alpha +\frac 14C_{\underline{\alpha }%
\underline{\beta }\underline{\gamma }}\left( u\right) ~e_\alpha ^{\underline{%
\alpha }}\left( u\right) \gamma ^{\underline{\beta }}\gamma ^{\underline{%
\gamma }}\right] \Psi ,  \label{covspinder}
\end{equation}
where the coefficients 
\[
C_{\underline{\alpha }\underline{\beta }\underline{\gamma }}\left( u\right)
=\left( D_\gamma e_{\underline{\alpha }}^\alpha \right) e_{\underline{\beta }%
\alpha }e_{\underline{\gamma }}^\gamma 
\]
are called the rotation Ricci coefficients; the covariant derivative $%
D_\gamma $ is defined by the usual Christoffel symbols for the off--diagonal
metric.

We can also define an equivalent covariant derivation of the Dirac spinor
field, $\overrightarrow{\nabla }_\alpha \Psi ,$ by using pentad
decompositions of the diagonalized d--metric (\ref{dmetric}),

\begin{equation}
\overrightarrow{\nabla }_\alpha \Psi =\left[ \delta _\alpha +\frac 14C_{%
\underline{\alpha }\underline{\beta }\underline{\gamma }}^{[\delta ]}\left(
u\right) ~f_\alpha ^{\underline{\alpha }}\left( u\right) \gamma ^{\underline{%
\beta }}\gamma ^{\underline{\gamma }}\right] \Psi ,  \label{covspindder}
\end{equation}
where there are introduced N--elongated partial derivatives and the
coefficients 
\[
C_{\underline{\alpha }\underline{\beta }\underline{\gamma }}^{[\delta
]}\left( u\right) =\left( D_\gamma ^{[\delta ]}f_{\underline{\alpha }%
}^\alpha \right) f_{\underline{\beta }\alpha }f_{\underline{\gamma }}^\gamma 
\]
are transformed into rotation Ricci d--coefficients which together with the
d--covariant derivative $D_\gamma ^{[\delta ]}$ are defined by anholonomic
pentads and anholonomic transforms of the Christoffel symbols.

For diagonal d--metrics the funfbein coefficients can be taken in their turn
in diagonal form and the corresponding gamma matrix $\widehat{\gamma }%
^\alpha \left( u\right) $ for anisotropic curved spaces are proportional to
the usual gamma matrix in flat spaces $\gamma ^{\underline{\gamma }}$. The
Dirac equations for locally anisotropic spacetimes are written in the
simplest form with respect to anholonomic frames,

\begin{equation}
(i\widehat{\gamma }^\alpha \left( u\right) \overrightarrow{\nabla _\alpha }%
-\mu )\Psi =0,  \label{diraceq}
\end{equation}
where $\mu $ is the mass constant of the Dirac field. The Dirac equations
are the Euler equations for the Lagrangian 
\begin{eqnarray}
&{\cal L}^{(1/2)}\left( u\right) &= \sqrt{\left| g\right| }\{[\Psi
^{+}\left( u\right) \widehat{\gamma }^\alpha \left( u\right) \overrightarrow{%
\nabla _\alpha }\Psi \left( u\right)  \label{direq} \\
&{}&-(\overrightarrow{\nabla _\alpha }\Psi ^{+}\left( u\right) )\widehat{%
\gamma }^\alpha \left( u\right) \Psi \left( u\right) ]-\mu \Psi ^{+}\left(
u\right) \Psi \left( u\right) \},  \nonumber
\end{eqnarray}
where by $\Psi ^{+}\left( u\right) $ we denote the complex conjugation and
transposition of the column$~\Psi \left( u\right) .$

Varying ${\cal L}^{(1/2)}$ on d--metric (\ref{direq}) we obtain the
symmetric energy--momentum d--tensor 
\begin{eqnarray}
\Upsilon _{\alpha \beta }\left( u\right) &=& \frac i4[\Psi ^{+}\left(
u\right) \widehat{\gamma }_\alpha \left( u\right) \overrightarrow{\nabla
_\beta }\Psi \left( u\right)  \nonumber \\
&{}& +\Psi ^{+}\left( u\right) \widehat{\gamma }_\beta \left( u\right) 
\overrightarrow{\nabla _\alpha }\Psi \left( u\right)  \nonumber \\
&{}&-(\overrightarrow{\nabla _\alpha }\Psi ^{+}\left( u\right) )\widehat{%
\gamma }_\beta \left( u\right) \Psi \left( u\right)  \nonumber \\
&{}& - (\overrightarrow{\nabla _\beta }\Psi ^{+}\left( u\right) )\widehat{%
\gamma }_\alpha \left( u\right) \Psi \left( u\right) ].  \label{diracemd}
\end{eqnarray}
We choose such spinor field configurations in curved spacetime as to be
satisfied the conditions (\ref{spinorem}).

One can introduce similar formulas to (\ref{diraceq})--(\ref{diracemd}) for
spacetimes provided with off-diagonal metrics with respect to holonomic
frames by changing of operators $\widehat{\gamma }_\alpha \left( u\right)
\rightarrow \gamma _\alpha \left( u\right) $ and $\overrightarrow{\nabla
_\beta }\rightarrow \nabla _\beta .$

\section{Anisotropic Taub NUT -- Dirac Spinor Solutions}

By straightforward calculations we can verify that because the conditions $%
D_\gamma ^{[\delta ]}f_{\underline{\alpha }}^\alpha =0$ are satisfied the
Ricci rotation coefficients vanishes, 
\[
C_{\underline{\alpha }\underline{\beta }\underline{\gamma }}^{[\delta
]}\left( u\right) =0\mbox{ and }\overrightarrow{\nabla _\alpha }\Psi =\delta
_\alpha \Psi ,
\]
and the anisotropic Dirac equations (\ref{diraceq}) transform into 
\begin{equation}
(i\widehat{\gamma }^\alpha \left( u\right) \delta _\alpha -\mu )\Psi =0.
\label{diraceq1}
\end{equation}

Further simplifications are possible for Dirac fields depending only on
coordinates $(t,x^1=r,x^2=\theta )$, i. e. $\Psi =\Psi (x^k)$ when the
equation (\ref{diraceq1}) transforms into 
\[
(i\gamma ^{\underline{0}}\partial _t+i\gamma ^{\underline{1}}\frac 1{\sqrt{%
\left| g_1\right| }}\partial _1+i\gamma ^{\underline{2}}\frac 1{\sqrt{\left|
g_2\right| }}\partial _2-\mu )\Psi =0.
\]
The equation (\ref{diraceq1}) simplifies substantially in $\zeta $%
--coordinates 
\[
\left( t,\zeta ^1=\zeta ^1(r,\theta ),\zeta ^2=\zeta ^2(r,\theta )\right) ,
\]
defined as to be satisfied the conditions 
\begin{equation}
\frac \partial {\partial \zeta ^1}=\frac 1{\sqrt{\left| g_1\right| }}%
\partial _1\mbox{ and }\frac \partial {\partial \zeta ^2}=\frac 1{\sqrt{%
\left| g_2\right| }}\partial _2  \label{zetacoord}
\end{equation}
We get 
\begin{equation}
(-i\gamma _{\underline{0}}\frac \partial {\partial t}+i\gamma _{\underline{1}%
}\frac \partial {\partial \zeta ^1}+i\gamma _{\underline{2}}\frac \partial {%
\partial \zeta ^2}-\mu )\Psi (t,\zeta ^1,\zeta ^2)=0.  \label{diraceq2}
\end{equation}
The equation (\ref{diraceq2}) describes the wave function of a Dirac
particle of mass $\mu $ propagating in a three dimensional Minkowski flat
plane which is imbedded as an anisotropic distribution into a 5D
pseudo--Riemannian spacetime.

The solution $\Psi = \Psi (t,\zeta ^1,\zeta ^2)$ of (\ref{diraceq2}) can be
written 
\[
\Psi =\left\{ 
\begin{array}{rcl}
\Psi ^{(+)}(\zeta ) & = & \exp {[-i(k_0t+k_1\zeta ^1+k_2\zeta ^2)]}\varphi
^0(k) \\ 
&  & \mbox{for positive energy;} \\ 
\Psi ^{(-)}(\zeta ) & = & \exp {[i(k_0t+k_1\zeta ^1+k_2\zeta ^2)]}\chi ^0(k)
\\ 
&  & \mbox{for negative energy,}
\end{array}
\right. 
\]
with the condition that $k_0$ is identified with the positive energy and $%
\varphi ^0(k)$ and $\chi ^0(k)$ are constant bispinors. To satisfy the
Klein--Gordon equation we must have 
\[
k^2={k_0^2-k_1^2-k_2^2}=\mu ^2. 
\]
The Dirac equations implies 
\[
(\sigma ^ik_i-\mu )\varphi ^0(k)\mbox{ and }(\sigma ^ik_i+\mu )\chi ^0(k), 
\]
where $\sigma ^i(i=0,1,2)$ are Pauli matrices corresponding to a realization
of gamma matrices as to a form of splitting to usual Pauli equations for the
bispinors $\varphi ^0(k)$ and $\chi ^0(k).$

In the rest frame for the horizontal plane parametrized by coordinates $%
\zeta =\{t,\zeta ^1,\zeta ^2\}$ there are four independent solutions of the
Dirac equations, 
\[
\varphi _{(1)}^0(\mu ,0)=\left( 
\begin{array}{c}
1 \\ 
0 \\ 
0 \\ 
0
\end{array}
\right) ,\ \varphi _{(2)}^0(\mu ,0)=\left( 
\begin{array}{c}
0 \\ 
1 \\ 
0 \\ 
0
\end{array}
\right) ,\ 
\]
\[
\chi _{(1)}^0(\mu ,0)=\left( 
\begin{array}{c}
0 \\ 
0 \\ 
1 \\ 
0
\end{array}
\right) ,\ \chi _{(2)}^0(\mu ,0)=\left( 
\begin{array}{c}
0 \\ 
0 \\ 
0 \\ 
1
\end{array}
\right) .
\]
In order to satisfy the conditions (\ref{spinorem}) for compatibility of the
equations (\ref{einsteq3a})--(\ref{einsteq3d}) we must consider wave packets
of type (for simplicity, we can use only superpositions of positive energy
solutions) 
\begin{eqnarray}
\Psi ^{(+)}(\zeta ) &=&\int \frac{d^3p}{2\pi ^3}\frac \mu {\sqrt{\mu
^2+(k^2)^2}}  \nonumber \\
&{}&\times \sum_{[\alpha ]=1,2,3}b(p,[\alpha ])\varphi ^{[\alpha ]}(k)\exp {%
[-ik_i\zeta ^i]}  \label{packet}
\end{eqnarray}
when the coefficients $b(p,[\alpha ])$ define a current (the group velocity) 
\begin{eqnarray}
&&J^2=\sum_{[\alpha ]=1,2,3}\int \frac{d^3p}{2\pi ^3}\frac \mu {\sqrt{\mu
^2+(k^2)^2}}|b(p,[\alpha ])|^2\frac{p^2}{\sqrt{\mu ^2+(k^2)^2}}  \nonumber \\
&&{}\equiv <\frac{p^2}{\sqrt{\mu ^2+(k^2)^2}}>  \nonumber
\end{eqnarray}
with $|p^2|\sim \mu $ and the energy--momentum d--tensor (\ref{diracemd})
has the next nonrivial coefficients 
\begin{eqnarray}
\Upsilon _0^0 &=&2\Upsilon (\zeta ^1,\zeta ^2)=k_0\Psi ^{+}\gamma _0\Psi , 
\nonumber \\
\Upsilon _1^1 &=&-k_1\Psi ^{+}\gamma _1\Psi ,\Upsilon _2^2=-k_2\Psi
^{+}\gamma _2\Psi \   \label{compat}
\end{eqnarray}
where the holonomic coordinates can be reexpressed $\zeta ^i=\zeta ^i(x^i).$
We must take two or more waves in the packet and choose such coefficients $%
b(p,[\alpha ]),$ satisfying corresponding algebraic equations, as to have in
(\ref{compat}) the equalities 
\begin{equation}
\Upsilon _1^1=\Upsilon _2^2=\Upsilon (\zeta ^1,\zeta ^2)=\Upsilon (x^1,x^2),
\label{compat1}
\end{equation}
required by the conditions (\ref{diracemd}).

\section{Taub NUT Solutions with Generic Local Anisotropy}

The Kaluza-Klein monopole \cite{GP,So} was obtained by embedding the
Taub-NUT gravitational instanton into five-dimensional theory, adding the
time coordinate in a trivial way. There are anisotropic variants of such
solutions \cite{vvt} when anisotropies are modelled by effective
polarizations of the induced magnetic field. The aim of this Section is to
analyze such Taub--NUT solutions for both cases of locally isotropic and
locally anisotropic configurations.

\subsection{A conformal transform of the Taub NUT metric}

We consider the Taub NUT solutions and introduce a conformal transformation
and a such redefinition of variables which will be useful for further
generalizations to anisotropic vacuum solutions.

\subsubsection{The Taub NUT solution}

This locally isotropic solution of the 5D vacuum Einstein equations is
expressed by the line element 
\begin{eqnarray}
ds_{(5D)}^2 &=&dt^2+ds_{(4D)}^2;  \label{nut} \\
ds_{(4D)}^2 &=&-V^{-1}(dr^2+r^2d\theta ^2 +\sin ^2\theta d\varphi^2) 
\nonumber \\
&{}& -V(dx^4+A_idx^i)^2\,  \nonumber
\end{eqnarray}
where 
\[
V^{-1}=1+\frac{m_0}r,m_0=const. 
\]
The functions $A_i$ are static ones associated to the electromagnetic
potential, 
\[
A_r=0,A_\theta =0,A_\varphi =4m_0\left( 1-\cos \theta \right) 
\]
resulting into ''pure'' magnetic field 
\begin{equation}
\vec{B}\,={\rm rot}\,\vec{A}=m_0\frac{\overrightarrow{r}}{r^3}\,.
\label{magnetic}
\end{equation}
of a Euclidean instanton; $\overrightarrow{r}$ is the spherical coordinate's
unity vector. The spacetime defined by (\ref{nut}) has the {\em global}
symmetry of the group $G_s=SO(3)\otimes U_4(1)\otimes T_t(1)$ since the line
element is invariant under the global rotations of the Cartesian space
coordinates and $y^4$ and $t$ translations of the Abelian groups $U_4(1)$
and $T_t(1)$ respectively. We note that the $U_4(1)$ symmetry eliminates the
so called NUT singularity if $y^4$ has the period $4\pi m_0$.

\subsubsection{Conformally transformed Taub NUT metrics}

With the aim to construct anisotropic generalizations it is more convenient
to introduce a new 5th coordinate, 
\begin{equation}
y^4\rightarrow \varsigma =y^4-\int \mu ^{-1}(\theta ,\varphi )d\xi (\theta
,\varphi ),  \label{coordch}
\end{equation}
with the property that 
\[
d\varsigma +4m_0(1-\cos \theta )d\theta =dy^4+4m_0(1-\cos \theta )d\varphi , 
\]
which holds for 
\[
d\xi =\mu (\theta ,\varphi )d(\varsigma -y^4)=\frac{\partial \xi }{\partial
\theta }d\theta +\frac{\partial \xi }{\partial \varphi }d\varphi , 
\]
when 
\begin{eqnarray*}
\frac{\partial \xi }{\partial \theta } &=&4m_0(1-\cos \theta )\mu , \\
\frac{\partial \xi }{\partial \varphi } &=&-4m_0(1-\cos \theta )\mu ,
\end{eqnarray*}
and, for instance, 
\[
\mu =\left( 1-\cos \theta \right) ^{-2}\exp [\theta -\varphi ]. 
\]
The changing of coordinate (\ref{coordch}) describe a reorientation of the
5th coordinate in a such way as we could have only one nonvanishing
component of the electromagnetic potential 
\[
A_\theta =4m_0\left( 1-\cos \theta \right) . 
\]

The next step is to perform a conformal transform, 
\[
ds_{(4D)}^2\rightarrow d\widehat{s}_{(4D)}^2=Vds_{(4D)}^2 
\]
and to consider the 5D metric

\begin{eqnarray}
ds_{(5D)}^2 &=&dt^2+d\widehat{s}_{(4D)}^2;  \label{conf4d} \\
d\widehat{s}_{(4D)}^2 &=&-(dr^2+r^2d\theta ^2)- r^2\sin ^2\theta d\varphi ^2
\nonumber \\
&{} & -V^2(d\zeta +A_\theta d\theta )^2,\,  \nonumber
\end{eqnarray}
(not being an exact solution of the Einstein equations) which will transform
into some exact solutions after corresponding anholonomic transforms.

Here, we emphasize that we chose the variant of transformation of a locally
isotropic non--Einsteinian metrics into an anisotropic one solving the
vacuum Einstein equations in order to illustrate a more simple procedure of
construction of 5D vacuum metrics with generic local anisotropy. As a metter
of principle we could remove vacuum isotropic solutions into vacuum
anisotropic ones, but the formula in this case would became very
combersome.. The fact of selection as an isotropic 4D Riemannian background
just the metric from the linear interval $d\widehat{s}_{(4D)}^2$ can be
treated as a conformal transformation of an instanton solution which is
anisotropically deformed and put trivially (by extension to the time like
coordinate) into a 5D metric as to generate a locally isotropic vacuum
gravitational field.

\subsection{Anisotropic Taub NUT solutions with mag\-netic
polarization}

We outline two classes of exact solutions of 5D vacuum Einstein equations
with generic anisotropies (see details in Ref. \cite{vvt}) which will be
extended to configurations with spinor matter field source.

\subsubsection{Solutions with angular polarization}

The ansatz for a d--metric (\ref{dmetric}), with a distinguished
anisot\-rop\-ic dependence on the angular coordinate $\varphi ,$ when $%
s=\varphi ,$ is taken in the form 
\begin{eqnarray*}
\delta s^2 &=&dt^2-\delta s_{(4D)}^2, \\
\delta s_{(4D)}^2 &=&-(dr^2+r^2d\theta ^2)- r^2\sin ^2\theta d\varphi ^2 
\nonumber \\
&{}& -V^2(r)\eta _4^2(\theta ,\varphi )\delta \varsigma ^2, \\
\delta \varsigma &=&d\varsigma +n_2(\theta ,\varphi )d\theta ,
\end{eqnarray*}
where the values $\eta _4^2(\theta ,\varphi )$ (we use non--negative values $%
\eta _4^2$ not changing the signature of metrics) and $n_2(\theta ,\varphi )$
must be found as to satisfy the vacuum Einstein equations in the form (\ref
{einsteq3a})--(\ref{einsteq3d}). We can verify that the data 
\begin{eqnarray}
x^0 &=&t,x^1=r,x^2=\theta ,y^3=s=\varphi ,y^4=\varsigma ,  \label{sol1} \\
g_0 &=&1,g_1=-1,g_2=-r^2,h_3=-r^2\sin ^2\theta ,  \nonumber \\
h_4 &=&V^2\left( r\right) \eta _{(\varphi )}^2,\eta _{(\varphi
)}^2=[1+\varpi (r,\theta )\varphi ]^2,w_i=0;  \nonumber \\
n_{0,1} &=&0; n_2 =n_{2[0]}\left( r,\theta \right) + n_{2[1]}\left( r,\theta
\right) /[1+\varpi (r,\theta )\varphi ]^2.  \nonumber
\end{eqnarray}
give an exact solution. If we impose the condition to obtain in the locally
isotropic limit just the metric (\ref{conf4d}), we have to choose the
arbitrary functions from the general solution of (\ref{einsteq3b}) as to
have 
\[
\eta _{(\varphi )}^2=[1+\varpi (r,\theta )\varphi ]^2\rightarrow 1%
\mbox{ for
}\varpi (r,\theta )\varphi \rightarrow 0. 
\]
For simplicity, we can analyze only angular anisotropies with $\varpi
=\varpi (\theta ),$ when 
\[
\eta _{(\varphi )}^2=\eta _{(\varphi )}^2(\theta ,\varphi )=[1+\varpi
(\theta )\varphi ]^2. 
\]

In the locally isotropic limit of the solution for $n_2\left( r,\theta
,\varphi \right) $, when $\varpi \varphi \rightarrow 0,$ we could obtain the
particular magnetic configuration contained in the metric (\ref{conf4d}) if
we impose the condition that 
\[
n_{2[0]}\left( r,\theta \right) +n_{2[1]}\left( r,\theta \right) =A_\theta
=4m_0\left( 1-\cos \theta \right) , 
\]
which defines only one function from two unknown values $n_{2[0]}\left(
r,\theta \right) $ and $n_{2[1]}\left( r,\theta \right) .$ This could have a
corresponding physical motivation. From the usual Kaluza--Klein procedure we
induce the 4D gravitational field (metric) and 4D electromagnetic field
(potentials $A_i),$ which satisfy the Maxwell equations in 4D
pseudo--Riemannian spacetime. For the case of spherical, locally isotropic,
symmetries the Maxwell equations can be written for vacuum magnetic fields
without any polarizations. When we introduce into consideration anholonomic
constraints and locally anisotropic gravitational configurations the
effective magnetic field could be effectively renormalized by higher
dimension gravitational field. This effect, for some classes of
anisotropies, can be modeled by considering that the constant $m_0$ is
polarized, 
\[
m_0\rightarrow m\left( r,\theta ,\varphi \right) =m_0\eta _m\left( r,\theta
,\varphi \right) 
\]
for the electromagnetic potential and resulting magnetic field. For ''pure"
angular anisotropies we write that 
\begin{eqnarray}
n_2\left( \theta ,\varphi \right)& = &n_{2[0]}\left( \theta \right)
+n_{2[1]}\left( \theta \right) /[1+\varpi (\theta )\varphi ]^2  \nonumber \\
& {} & =4m_0\eta _m\left( \theta ,\varphi \right) \left( 1-\cos \theta
\right) ,  \nonumber
\end{eqnarray}
for 
\[
\eta _{(\varphi )}^2\left( \theta ,\varphi \right) =\eta _{(\varphi
)[0]}^2\left( \theta \right) +\eta _{(\varphi )[1]}^2\left( \theta \right)
/[1+\varpi (\theta )\varphi ]^2. 
\]
This could result in a constant angular renormalization even $\varpi (\theta
)\varphi \rightarrow 0.$

\subsubsection{Solutions with extra--dimension induced polarization}

Another class of solutions is constructed if we consider a d--metric of the
type (\ref{dmetric}), when $s=\varsigma ,$ with anisotropic dependence on
the 5th coordinate $\varsigma ,$ 
\begin{eqnarray*}
\delta s^2 &=&dt^2-\delta s_{(4D)}^2, \\
\delta s_{(4D)}^2 &=&-(dr^2+r^2d\theta ^2)-r^2\sin ^2\theta d\varphi ^2 \\
&{}&-V^2(r)\eta _{(\varsigma )}^2(\theta ,\varsigma )\delta \varsigma ^2, \\
\delta \varsigma  &=&d\varsigma +w_3(\theta ,\varsigma )d\theta ,
\end{eqnarray*}
where, for simplicity, we omit possible anisotropies on variable $r,$ i. e.
we state that $\eta _{(\varsigma )}$ and $w_2$ are not functions of $r.$

The data for a such solution are 
\begin{eqnarray}
x^0 &=&t,x^1=r,x^2=\theta ,y^3=s=\varsigma ,y^4=\varphi ,  \label{sol2} \\
g_0 &=&1,g_1=-1,g_2=-r^2,h_4=-r^2\sin ^2\theta ,  \nonumber \\
h_3 &=&V^2\left( r\right) \eta _{(\varsigma )}^2,\eta _{(\varsigma )}^2=\eta
_{(\varsigma )}^2(r,\theta ,\varsigma ),n_{0,1}=0;  \nonumber \\
w_{0,1} &=&0,w_2=4m_0\eta _m\left( \theta ,\varsigma \right) \left( 1-\cos
\theta \right) ,n_0=0,  \nonumber \\
n_{1,2} &=&n_{1,2[0]}\left( r,\theta \right) +n_{1,2[1]}\left( r,\theta
\right) \int \eta _{(\varsigma )}^{-3}(r,\theta ,\varsigma )d\varsigma , 
\nonumber
\end{eqnarray}
where the function $\eta _{(\varsigma )}=\eta _{(\varsigma )}(r,\theta
,\varsigma )$ is an arbitrary one as follow for the case $h_4^{*}=0,$ for
angular polarizations we state, for simplicity, that $\eta _{(\varsigma )}$
does not depend on $r,$ i. e. $\eta _{(\varsigma )}=\eta _{(\varsigma
)}(\theta ,\varsigma ).$ We chose the coefficient 
\[
w_4=4m_0\eta _m\left( \theta ,\varsigma \right) \left( 1-\cos \theta \right) 
\]
as to have compatibility with the locally isotropic limit when $w_2\simeq
A_\theta $ with a ''polarization'' effect modeled by $\eta _m\left( \theta
,\varsigma \right) ,$ which could have a constant component $\eta _m\simeq
\eta _{m[0]}=const$ for small anisotropies. In the simplest cases we can fix
the conditions $n_{1,2[0,1]}\left( r,\theta \right) =0.$ All functions $\eta
_{(\varsigma )}^2,\eta _m$ and $n_{1,2[0,1]}$ can be treated as some
possible induced higher dimensional polarizations.

\section{Anisotropic Taub NUT--Dirac Fields}

In this Section we construct two new classes of solutions of the 5D
Einstein--Dirac fields in a manner as to extend the locally anisotropic Taub
NUT metrics defined by data (\ref{sol1}) and (\ref{sol2}) as to be solutions
of the Einstein equations (\ref{einsteq3a})--(\ref{einsteq3d}) with a
nonvanishing diagonal energy momentum d--tensor 
\[
\Upsilon _\beta ^\alpha =\{2\Upsilon (r,\theta ),\Upsilon (r,\theta
),\Upsilon (r,\theta ),0,0\} 
\]
for a Dirac wave packet satisfying the conditions (\ref{compat}) and (\ref
{compat1}).

\subsection{Dirac fields and angular polarizations}

In order to generate from the data (\ref{sol1}) a new solution with Dirac
spinor matter field we consider instead of a linear dependence of
polarization, $\eta _{(\varphi )}\sim [1+\varpi \left( r,\theta \right)
\varphi ],$ an arbitrary function $\eta _{(\varphi )}\left( r,\theta
,\varphi \right) $ for which $h_4=V^2(r)\eta _{(\varphi )}^2\left( r,\theta
,\varphi \right) $ is an exact solution of the equation (\ref{einsteq3b})
with $\Upsilon _1=\Upsilon \left( r,\theta \right) .$ With respect to the
variable $\eta _{(\varphi )}^2\left( r,\theta ,\varphi \right) $ this
component of the Einstein equations becomes linear 
\begin{equation}
\eta _{(\varphi )}^{**}+r^2\sin ^2\theta \Upsilon \eta _{(\varphi )}=0
\label{eqaux11}
\end{equation}
which is a second order linear differential equation on variable $\varphi $
with parametric dependencies of the coefficient $r^2\sin ^2\theta \Upsilon $
on coordinates $\left( r,\theta \right) .$ The solution of equation (\ref
{eqaux11}) is to be found following the method outlined in Ref. \cite{kamke}%
: 
\begin{eqnarray}
\eta _{(\varphi )} &=&C_1\left( r,\theta \right) \cosh [\varphi r\sin \theta 
\sqrt{\left| \Upsilon \left( r,\theta \right) \right| }+C_2\left( r,\theta
\right) ],  \nonumber \\
&{}&\Upsilon \left( r,\theta \right) <0;  \label{solaux21} \\
&=&C_1\left( r,\theta \right) +C_2\left( r,\theta \right) \varphi ,\Upsilon
\left( r,\theta \right) =0;  \label{solaux22} \\
&=&C_1\left( r,\theta \right) \cos [\varphi r\sin \theta \sqrt{\Upsilon
\left( r,\theta \right) }+C_2\left( r,\theta \right) ],  \nonumber \\
&{}&\Upsilon \left( r,\theta \right) >0,  \label{solaux23}
\end{eqnarray}
where $C_{1,2}\left( r,\theta \right) $ are some functions to be defined
from some boundary conditions. The first solution (\ref{solaux21}), for
negative densities of energy should be excluded as unphysical, the second
solution (\ref{solaux22}) is just that from (\ref{sol1}) for the vacuum
case. A new interesting physical situation is described by the solution (\ref
{solaux23}) when we obtain a Taub NUT anisotropic metric with periodic
anisotropic dependencies on the angle $\varphi $ where the periodicity could
variate on coordinates $\left( r,\theta \right) $ as it is defined by the
energy density $\Upsilon \left( r,\theta \right) .$ For simplicity, we can
consider a package of spinor waves with constant value of $\Upsilon
=\Upsilon _0$ and fix some boundary and coordinate conditions when $%
C_{1,2}=C_{1,2[0]}$ are constant. This type of anisotropic Taub NUT
solutions are described by a d--metric coefficient 
\begin{equation}
h_4=V^2(r)C_{1[0]}^2\cos ^2[\varphi r\sin \theta \sqrt{\Upsilon _0}%
+C_{2[0]}].  \label{aux31}
\end{equation}
Putting this value into the formulas (\ref{alpha1}), (\ref{alpha2}) and (\ref
{beta}) for coefficients in equations (\ref{einsteq3c}) we can express $%
\alpha _{1,2}=\alpha _{1,2}[h_3,h_4,\Upsilon _0]$ and $\beta =\beta
[h_3,h_4,\Upsilon _0]$ (we omit these rather simple but cumbersome formulas)
and in consequence we can define the values $w_{1,2}$ by solving linear
algebraic equations: 
\[
w_{1,2}\left( r,\theta ,\varphi \right) =\alpha _{1,2}\left( r,\theta
,\varphi \right) /\beta \left( r,\theta ,\varphi \right) .
\]

Having defined the values (\ref{aux31}) it is a simple task of two
integrations on $\varphi $ in order to define 
\begin{eqnarray}
n_2 &=&n_{2[0]}\left( r,\theta \right) \left[ \ln \frac{1+\cos {\tilde{\kappa%
}}}{1-\cos {\tilde{\kappa}}}+\frac 1{1-\cos {\tilde{\kappa}}}+\frac 1{1-\sin 
{\tilde{\kappa}}}\right]   \nonumber \\
&{}&+n_{2[1]}\left( r,\theta \right) ,  \label{n2conf}
\end{eqnarray}
were 
\[
{\tilde{\kappa}}=\varphi r\sin \theta \sqrt{\Upsilon _0}+C_{2[0]},
\]
$n_{2[0,1]}\left( r,\theta \right) $ are some arbitrary functions to be
defined by boundary conditions. We put $n_{0,1}=0$ to obtain in the vacuum
limit the solution (\ref{sol1}).

Finally, we can summarize the data defining an exact solution for an
anisotropic (on angle $\varphi $) Dirac wave packet -- Taub NUT
configuration: 
\begin{eqnarray}
x^0 &=&t,x^1=r,x^2=\theta ,y^3=s=\varphi ,y^4=\varsigma ,  \label{sol1a} \\
g_0 &=&1,g_1=-1,g_2=-r^2,h_3=-r^2\sin ^2\theta ,  \nonumber \\
h_4 &=&V^2\left( r\right) \eta _{(\varphi )}^2,\eta _{(\varphi )}=C_1\left(
r,\theta \right) \cos {\tilde {\kappa}} (r,\theta ,\varphi ),  \nonumber \\
w_i &=&0,n_{0,1}=0,n_2=n_2\left( r,\theta ,{\tilde {\kappa}} (r,\theta
,\varphi )\right) \mbox{ see  (\ref{n2conf})},  \nonumber \\
\Psi &=&\Psi ^{(+)}\left( \zeta ^{1,2}(x^1,x^2)\right) 
\mbox{ see 
(\ref{packet})},  \nonumber \\
\Upsilon &=&\Upsilon \left( \zeta ^{1,2}(x^1,x^2)\right) 
\mbox{ see 
(\ref{compat})}.  \nonumber
\end{eqnarray}
This solution will be extended to additional soliton anisotropic
configurations in the next Section.

\subsection{Dirac fields and extra dimension polarizations}

Now we consider a generalization of the data (\ref{sol2}) for generation of
a new solution, with generic local anisotropy on extra dimension 5th
coordinate, of the Einstein -- Dirac equations. Following the equation (\ref
{einsteq3c}) we conclude that there are not nonvacuum solutions of the
Einstein equations (with $\Upsilon \neq 0)$ if $h_4^{*}=0$ which impose the
condition $\Upsilon =0$ for $h_3,h_4\neq 0.$ So, we have to consider that
the d--metric component $h_4=-r^2\sin ^2\theta $ from the data (\ref{sol2})
is generalized to a function $h_4\left( r,\theta ,\varsigma \right) $
satisfying a second order nonlinear differential equation on variable $%
\varsigma $ with coefficients depending parametrically on coordinates $%
\left( r,\theta \right) .$ The equation (i. e. (\ref{einsteq3c})) can be
linearized (see Ref. \cite{kamke}) if we introduce a new variable $h_4=h^2,$%
\[
h^{**}-\frac{h_3^{*}}{2h_3}h^{*}+h_3\Upsilon h=0,
\]
which, in its turn, can be transformed :

a) to a Riccati form if we introduce a new variable $v,$ for which $%
h=v^{*}/v,$%
\begin{equation}
v^{*}+v^2-\frac{h_3^{*}}{2h_3}v+h_3\Upsilon =0;  \label{riccati1}
\end{equation}

b) to the so--called normal form \cite{kamke}, 
\begin{equation}
\lambda ^{**}+I\lambda =0,  \label{normal}
\end{equation}
obtained by a redefinition of variables like 
\[
\lambda =h\exp \left[ -\frac 14\int \frac{h_3^{*}}{h_3}d\varsigma \right]
=h\ h_3^{-1/4}
\]
where 
\[
I=h_3\Upsilon -\frac 1{16}\frac{h_3^{*}}{h_3}+\frac 14\left( \frac{h_3^{*}}{%
h_3}\right) ^{*}.
\]
We can construct explicit series and/or numeric solutions (for instance, by
using Mathematica or Maple programs) of both type of equations (\ref
{riccati1}) and normal (\ref{normal}) for some stated boundary conditions
and type of polarization of the coefficient $h_3\left( r,\theta ,\varsigma
\right) =V^2\left( r\right) \eta _{(\varsigma )}^2(r,\theta ,\varsigma )$
and, in consequence, to construct different classes of solutions for $%
h_4\left( r,\theta ,\varsigma \right) .$ In order to have compatibility with
the data (\ref{sol2}) we must take $h_4$ in the form 
\[
h_4\left( r,\theta ,\varsigma \right) =-r^2\sin ^2\theta +h_{4(\varsigma
)}\left( r,\theta ,\varsigma \right) ,
\]
where $h_{4(\varsigma )}\left( r,\theta ,\varsigma \right) $ vanishes for $%
\Upsilon \rightarrow 0.$

Having defined a value of $h_4\left( r,\theta ,\varsigma \right) $ we can
compute the coefficients (\ref{alpha1}), (\ref{alpha2}) and (\ref{beta}) and
find from the equations (\ref{einsteq3c})

\[
w_{1,2}\left( r,\theta ,\varsigma \right) =\alpha _{1,2}\left( r,\theta
,\varsigma \right) /\beta \left( r,\theta ,\varsigma \right) . 
\]

From the equations (\ref{einsteq3d}), after two integrations on variable $%
\varsigma $ one obtains the values of $n_{1,2}\left( r,\theta ,\varsigma
\right) .$ Two integrations of equations (\ref{einsteq3d}) define 
\[
n_i(r,\theta ,\varsigma )=n_{i[0]}(r,\theta )\int_0^\varsigma
dz\int_0^zdsP(r,\theta ,s)+n_{i[1]}(r,\theta ), 
\]
where 
\[
P\equiv \frac 12(\frac{h_3^{*}}{h_3}-3\frac{h_4^{*}}{h_4}) 
\]
and the functions $n_{i[0]}(r,\theta )$ and $n_{i[1]}(r,\theta )$ on $%
(r,\theta )$ have to be defined by solving the Cauchy problem. The boundary
conditions of both type of coefficients $w_{1,2}$ and $n_{1,2}$ should be
expressed in some forms transforming into corresponding values for the data (%
\ref{sol2}) if the source $\Upsilon \rightarrow 0.$ We omit explicit
formulas for exact Einstein--Dirac solutions with $\varsigma $%
--polarizations because their forms depend very strongly on the type of
polarizations and vacuum solutions.

\section{Anholonomic Soliton--Taub NUT--Dirac Fi\-elds}

In the next subsections we analyze two explicit examples when the spinor
field induces two dimensional, depending on three variables, solitonic
anisotropies.

\subsection{Kadomtsev--Petviashvili type solitons}

By straightforward verification we conclude that the d--metric component $%
h_4(r,\theta ,s)$ could be a solution of Kadomtsev--Petviashvili (KdP)
equation \cite{kad} (the first methods of integration of 2+1 dimensional
soliton equations where developed by Dryuma \cite{dryuma} and Zakharov and
Shabat \cite{zakhsh}) 
\begin{equation}
h_4^{**}+\epsilon \left( \dot{h}_4+6h_4h_4^{\prime }+h_4^{\prime \prime
\prime }\right) ^{\prime }=0,\epsilon =\pm 1,  \label{kdp}
\end{equation}
if the component $h_3(r,\theta ,s)$ satisfies the Bernoulli equations \cite
{kamke} 
\begin{equation}
h_3^{*}+Y\left( r,\theta ,s\right) (h_3)^2+F_\epsilon \left( r,\theta
,s\right) h_3=0,  \label{bern1}
\end{equation}
where, for $h_4^{*}\neq 0,$%
\begin{equation}
Y\left( r,\theta ,s\right) =\kappa \Upsilon \frac{h_4}{h_4^{*}},
\label{sourse1}
\end{equation}
and 
\[
F_\epsilon \left( r,\theta ,s\right) =\frac{h_4^{*}}{h_4}+\frac{2\epsilon }{%
h_4^{*}}\left( \dot{h}_4+6h_4h_4^{\prime }+h_4^{\prime \prime \prime
}\right) ^{\prime }. 
\]
The three dimensional integral variety of (\ref{bern1}) is defined by
formulas 
\[
h_3^{-1}\left( r,\theta ,s\right) =h_{3(x)}^{-1}\left( r,\theta \right)
E_\epsilon \left( x^i,s\right) \times \int \frac{Y\left( r,\theta ,s\right) 
}{E_\epsilon \left( r,\theta ,s\right) }ds, 
\]
where 
\[
E_\epsilon \left( r,\theta ,s\right) =\exp \int F_\epsilon \left( r,\theta
,s\right) ds 
\]
and $h_{3(x)}\left( r,\theta \right) $ is a nonvanishing function.

In the vacuum case $Y\left( r,\theta ,s\right) =0$ and we can write the
integral variety of (\ref{bern1}) 
\[
h_3^{(vac)}\left( r,\theta ,s\right) =h_{3(x)}^{(vac)}\left( r,\theta
\right) \exp \left[ -\int F_\epsilon \left( r,\theta ,s\right) ds\right] . 
\]

We conclude that a solution of KdP equation (\ref{bern1}) could be generated
by a non--perturbative component $h_4(r,\theta ,s)$ of a diagonal h--metric
if the second component $h_3\left( r,\theta ,s\right) $ is a solution of
Bernoulli equations (\ref{bern1}) with coefficients determined both by $h_4$
and its partial derivatives and by the $\Upsilon _1^1$ component of the
energy--momentum d--tensor (see (\ref{compat1})). The parameters
(coefficients) of (2+1) dimensional KdP solitons are induced by gravity and
spinor constants and spinor waves coefficients defining locally anisotropic
interactions of packets of Dirac's spinor waves.

\subsection{(2+1) sine--Gordon type solitons}

In a similar manner we can prove that solutions $h_4(r,\theta ,s)$ of (2+1)
sine--Gordon equation (see, for instance, \cite{har,lieb,whith}) 
\[
h_4^{**}+h_4^{^{\prime \prime }}-\ddot{h}_4=\sin (h_4)
\]
also induce solutions for $h_3\left( r,\theta ,s\right) $ following from the
Bernoulli equation 
\[
h_3^{*}+\kappa E(r,\theta )\frac{h_4}{h_4^{*}}(h_3)^2+F\left( r,\theta
,s\right) h_3=0,h_4^{*}\neq 0,
\]
where 
\[
F\left( r,\theta ,s\right) =\frac{h_4^{*}}{h_4}+\frac 2{h_4^{*}}\left[
h_4^{^{\prime \prime }}-\ddot{h}_4-\sin (h_4)\right] .
\]
The general solutions (with energy--momentum sources and in vacuum cases)
are constructed by a corresponding redefinition of coefficients in the
formulas from the previous subsection. We note that we can consider both
type of anisotropic solitonic polarizations, depending on angular variable $%
\varphi $ or on extra dimension coordinate $\varsigma .$ Such classes of
solutions of the Einstein--Dirac equations describe three dimensional spinor
wave packets induced and moving self--consistently on solitonic
gravitational locally anisotropic configurations. In a similar manner, we
can consider Dirac wave packets generating and propagating on locally
anisotropic black hole (with rotation ellipsoid horizons), black tori,
anisotropic disk and two or three dimensional black hole anisotropic
gravitational structures \cite{vsol}. Finally, we note that such
gravitational solitons are induced by Dirac field matter sources and are
different from those soliton solutions of vacuum Einstein equations
originally considered by Belinski and Zakharov \cite{belinski}.

\section{ Conclusions}

We have argued that the anholonomic frame method can be applied for
construction on new classes of Einstein--Dirac equations in five dimensional
(5D) spacetimes. Subject to a form of metric ansatz with dependencies of
coefficients on two holonomic and one anholonomic variables we obtained a
very simplified form of field equations which admit exact solutions. We have
identified two classes of solutions describing Taub NUT like metrics with
anisotropic dependencies on angular parameter or on the fifth coordinate. We
have shown that both  classes of anisotropic vacuum solutions can be
generalized to matter sources with the energy--momentum tensor defined by
some wave packets of Dirac fields. Although the Dirac equation is a quantum
one, in the quasi--classical approximation we can consider such spinor
fields as some spinor waves propagating in a three dimensional Minkowski
plane which is imbedded in a self--consistent manner in a Taub--NUT
anisotropic spacetime. At the classical level it should be emphasized that
the results of this paper are very general in nature, depending in a crucial
way only on the locally Lorentzian nature of 5D spacetime and on the
supposition that this spacetime is constructed as a trivial time extension
of 4D spacetimes. We have proved that the new classes of solutions admit
generalizations to nontrivial topological configurations of 3D dimensional
solitons (induced by anisotropic spinor matter) defined as solutions
Kadomtsev--Petviashvili or sine--Gordon equations.


\subsection*{Acknowledgments}

The authors wish to thank M. Visinescu, E. Gaburov, D. Gontsa and O.
Tintareanu-Mircea for discussing the results and colaboration.
 The S. V. work was partially supported by "The 2000-2001 California
State University Legislative Award".


\end{document}